\documentclass[3p,times]{elsarticle}
\usepackage{ecrc}

\volume{00}

\firstpage{1}

\journalname{Annals of Physics}

\runauth{S. Longhi}

\usepackage{amssymb}
\usepackage[figuresright]{rotating}

\begin{document}

\begin{frontmatter}



\title{Quantum recurrence and fractional dynamic localization in ac-driven perfect state transfer Hamiltonians}


\author{Stefano Longhi}
\address{Dipartimento di Fisica, Politecnico di Milano and Istituto di Fotonica e Nanotecnologie del Consiglio Nazionale delle Ricerche, Piazza L. da Vinci 32, I-20133 Milano, Italy\\ Tel/Fax: 0039 022399 6156/6126, email: longhi@fisi.polimi.it}

\begin{abstract}
Quantum recurrence and dynamic localization are investigated in a class of ac-driven tight-binding Hamiltonians, the Krawtchouk quantum chain, which in the undriven case provides a paradigmatic Hamiltonian model that realizes perfect quantum state transfer and mirror inversion. The equivalence between the the ac-driven single-particle  Krawtchouk Hamiltonian $\hat{H}(t)$  and the non-interacting ac-driven bosonic junction Hamiltonian enables to determine in a closed form the quasi energy spectrum of $\hat{H}(t)$ and the conditions for exact wave packet reconstruction (dynamic localization). In particular, we show that quantum recurrence, which is predicted by the general quantum recurrence theorem, is {\it exact} for the  Krawtchouk quantum chain in a dense range of the driving amplitude. Exact quantum recurrence provides perfect wave packet reconstruction at a frequency which is {\it fractional} than the driving frequency, a phenomenon that can be referred to as fractional dynamic localization.
\end{abstract}


\end{frontmatter}

\section{Introduction}

Quantum mechanical spreading of a particle hopping on a tight binding lattice is known to be suppressed by the application of an external force, which leads to periodic wave packet reconstruction \cite{uff1,Dunlap}. The phenomenon of quantum  diffusion suppression and periodic wave packet relocalization induced by an ac force was originally predicted in a seminal paper by Dunlap and Kenkre \cite{Dunlap}, and since then it is referred to as dynamic localization (DL). This kind of
localization is conceptually very different from other
forms of localization, like Anderson localization 
or localization in the periodically kicked quantum rotator
problem, and arises from a collapse of the quasi energy spectrum at certain 'magic' driving amplitudes \cite{Holthaus}.
 DL was observed in different physical systems, including 
electronic transport in semiconductor superlattices \cite{semi}, 
 cold atoms and Bose-Einstein condensates in optical lattices \cite{BEC} and optical beams in curved waveguide arrays \cite{optics,optics2}.  
 Several works have extended the theory of DL to different physical conditions, for example to account for disorder \cite{disorder}, non-nearest neighbor hopping \cite{sterco}, 
 particle interaction \cite{inter}, and to non-Hermitian lattices \cite{LonghiNH}. 
 Lattice truncation is generally believed to be detrimental for perfect wave packet reconstruction. Indeed, quasi-energy level collapse becomes only approximate for lattices with a finite number of sites \cite{Holthaus,LonghiPRB2008,note}. In this case, even in the absence of the external ac field, approximate reconstruction should occur at long enough times according to the very general quantum recurrence theorem \cite{Loinger,periodic}, which is an extension to a quantum system with a discrete energy (or quasi energy) spectrum of the celebrated Pioncar\'{e} recurrence theorem of classical mechanics \cite{Poincare}.\\ 
 In this work we present an exactly-solvable ac-driven lattice model with a finite number of lattice sites in which perfect wave packet reconstruction can occur at a frequency which is {\it fractional} than the driving frequency, a phenomenon that we refer to as {\it fractional} DL. As opposed to ordinary DL, fractional DL corresponds  to a set of partial quasi-energy collapses and can occur at almost every amplitude of the driving force.  Fractional DL can be viewed as the realization of {\it exact}  quantum recurrence in a system with a discrete quasi-energy spectrum at finite times which are integer multiplies than the forcing oscillation period. The lattice model that we consider is the Krawtchouk quantum chain \cite{K0,K1,K2}, which has been extensively investigated  in several physical fields \cite{K0,K1,K2,K3,K4} for its property of realizing perfect quantum state transfer and mirror inversion in the undriven case \cite{K2,K3}. 

The paper is organized as follows. In Sec.2 we briefly introduce  the ac-driven Krawtchouk quantum chain and show its equivalence with the ac-driven double-well model for non-interacting bosons (ac-driven bosonic junction). A closed-form expressions of the quantum propagator and quasi energy spectrum  are derived in Sec.3, where the phenomena of quantum recurrence, fractional DL and mirror inversion are discussed. Numerical results of quasi energy spectra, perfect wave packet reconstruction and mirror imaging arising from fractional DL are presented in Sec.4. Finally, in Sec.5 the main conclusions are outlined.

\begin{figure}[b]
\includegraphics[width=10cm]{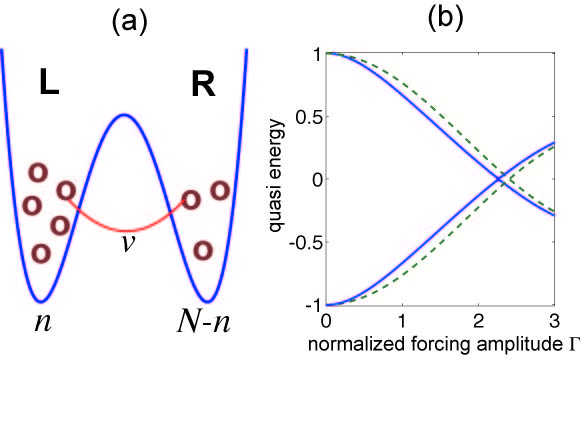}
\caption{(Color online). (a) Schematic of a bosonic junction. $N$ non-interacting bosons are trapped in a double well potential. The hopping rate between the two wells is $\nu$. An external force $F(t)$ is applied to the particles. (b)  Quasi energies $\pm \mu(\Gamma)$ of the sinusoidally-driven double-well [$F(t)=F_0 \cos (\omega t)$] versus the normalized forcing amplitude $\Gamma=F_0 / \omega$ for $\nu=1$ and $\omega=3$. The solid curves refer to the exact numerically-computed quasi energies, whereas the dashed curves are the approximate analytical curves defined by Eq.(23). Quasi energy crossing occurs at $\Gamma=\Gamma_1 \simeq 2.261$.} 
\end{figure}

\begin{figure}[b]
\includegraphics[width=16cm]{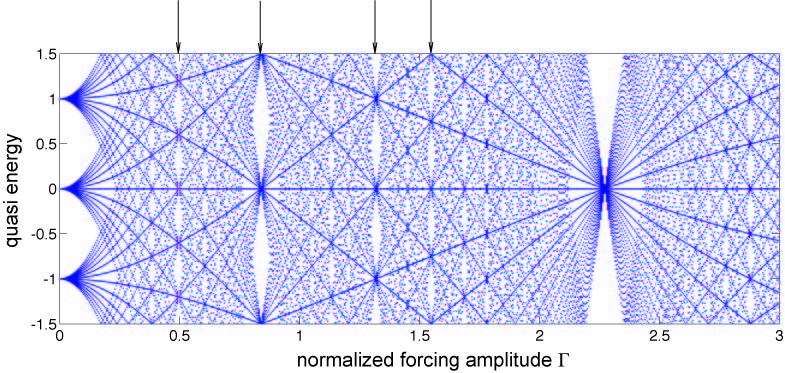}
\caption{(Color online). Quasi energy spectrum versus  normalized forcing amplitude $\Gamma=F_0 / \omega$ for he sinusoidally-driven Krawtchouk quantum chain with $(N+1)=41$ lattice sites for $\nu=1$ and $\omega=3$. The quasi energy is defined inside the range $(-\omega/2,\omega/2)$. The vertical arrows in the figure highlight partial collapse of quasi energy levels (fractional DL) at four values of $\Gamma$ (from left to right): 0.496, 0.84, 1.3172 and 1.5476.
Full band collapse, corresponding to ordinary DL,  occurs at $\Gamma=\Gamma_0 \simeq 2.261$.} 
\end{figure}

\begin{figure}[b]
\includegraphics[width=12cm]{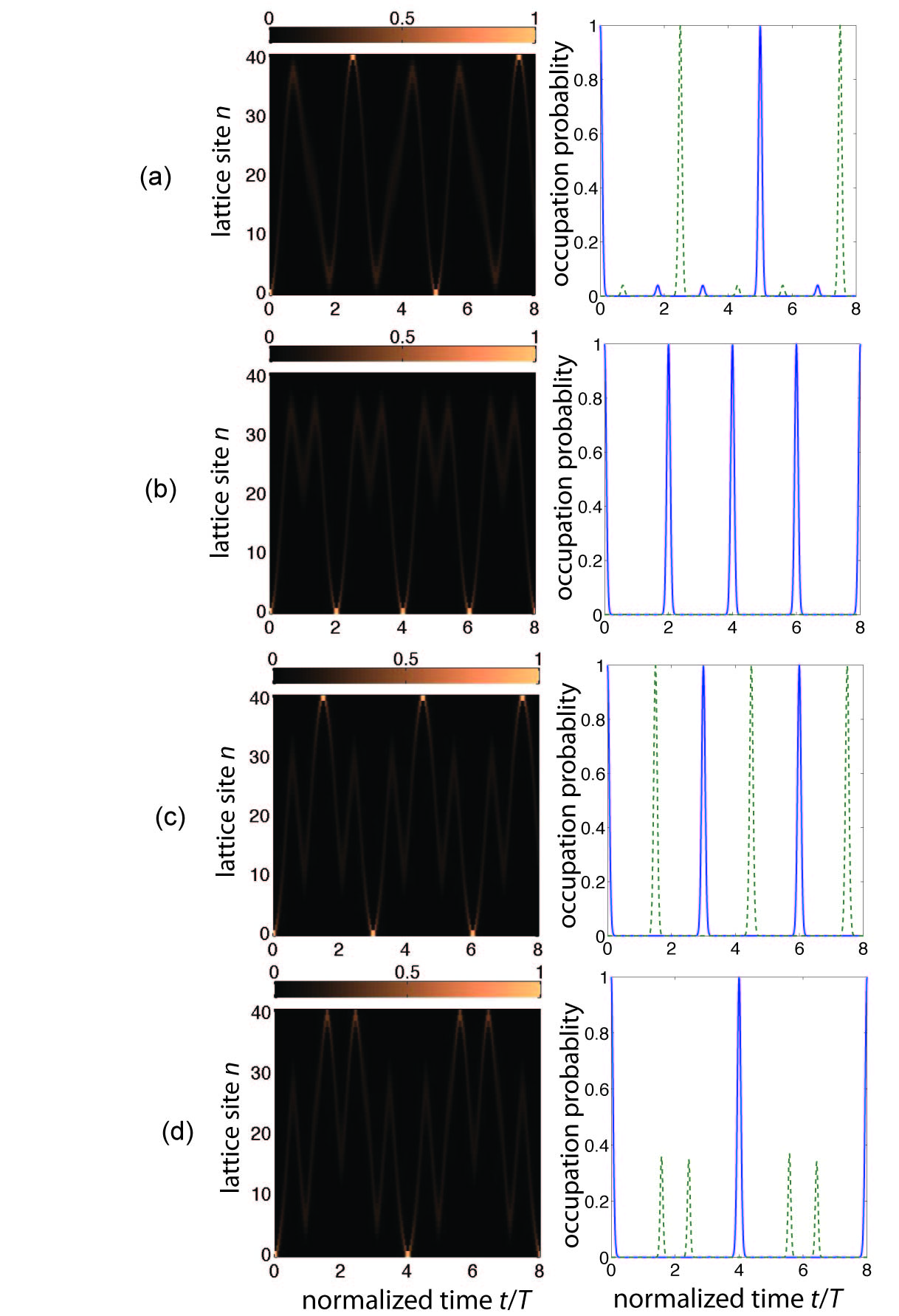}
\caption{(Color online). Numerically-computed evolution of the site occupation probabilities $|c_n(t)|^2$ for the sinusoidally-driven Krawtchouk quantum chain with $(N+1)=41$ lattice sites  for $\nu=1$, $\omega=3$,  and for a few values of the normalized forcing amplitude $\Gamma$ in the fractional dynamic localization regime: (a) 0.496, (b) 0.84, (c) 1.3172, and (d) 1.5476. The corresponding values of $2 \mu/ \omega=Q/M$ are 3/5, 1/2, 1/3 and 1/4, respectively. The initial condition is $c_{n}(0)=\delta_{n,0}$, corresponding to the particle initially occupying the left edge site of the chain. The left panels show in pseudo colors the evolution of $|c_n(t)|^2$ in the $(n,t)$ plane; the right panels show the detailed temporal evolution of the occupation probabilities of the  left- and right-oundary sites,  $|c_{0}(t)|^2$ (solid curves) and $|c_{N}(t)|^2$ (dashed curves). Time is normalized to the period $T= 2 \pi / \omega$ of the external force. Note that in all cases wave packet reconstruction is observed, whereas mirror imaging is attained only in (a) and (c), corresponding to odd values of $M$.} 
\end{figure}

\begin{figure}[b]
\includegraphics[width=12cm]{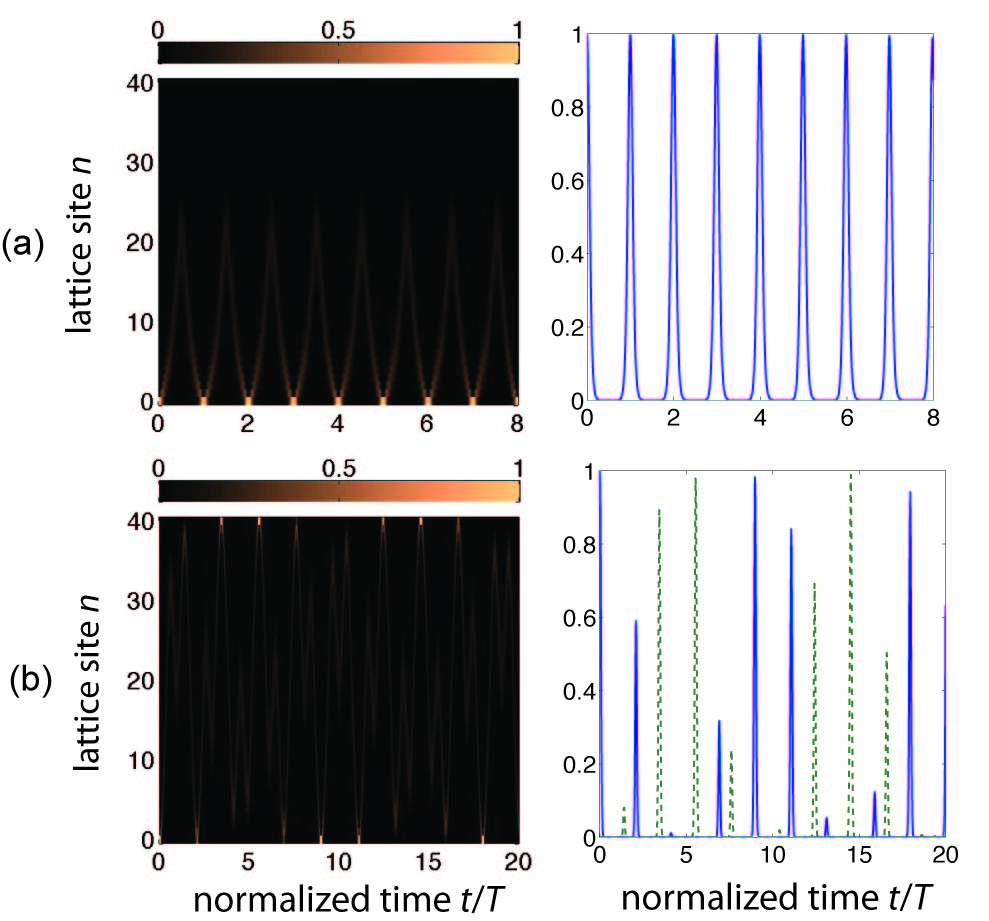}
\caption{(Color online). Same as Fig.3, but for  (a) $\Gamma=\Gamma_0=2.261$, corresponding to DL, and (b) $\Gamma=0.9965$, corresponding to a quasi-periodic dynamics [$ 2 \mu / \omega=1/ \sqrt{5}$].} 
\end{figure}

\section{ac-driven Krawtchouk quantum chain and non-interacting bosonic junction equivalence}
We consider the hopping motion of a quantum particle on a linear chain, composed by $(N+1)$ sites, driven by an external time-dependent force $F(t)$. In the nearest-neighbor tight-binding approximations, the quantum system is described by the time-dependent Hamiltonian
\begin{equation}
\hat{H}(t)= - \sum_{n=0}^{N}  \kappa_n \left(  | n+1 \rangle \langle n| + |n \rangle \langle n+1 |  \right) + F(t) \sum_{n=0}^{N} n |n \rangle \langle n |
\end{equation}
where $|n \rangle$ is the Wannier state that localizes the particle at lattice site $n$ and $\kappa_n$ is the hopping rate between adjacent sites $n$ and $(n+1)$. 
By expanding the particle  quantum state $| \psi(t) \rangle$ 
as a superposition of localized Wannier states $| \psi(t) \rangle=\sum_{n=0}^N c_n(t) |n \rangle$,  the evolution equations for the amplitude probabilities $c_n$ derived
from the Hamiltonian (1) (with $\hbar=1$) explicitly read
\begin{equation}
i \frac{dc_n}{dt}=-\kappa_n c_{n+1}-\kappa_{n-1}c_{n-1}+nF(t)c_n
\end{equation}
$(n=0,1,2,...,N$) with $\kappa_{-1}=\kappa_N=0$.
The Krawtchouk quantum chain corresponds to the following special sequence of hopping rates \cite{K0,K1,K2,K3}
\begin{equation}
\kappa_n= \nu \sqrt{(n+1)(N-n)}
\end{equation}
$(n=0,1,2,..., N$), where $\nu$ is a constant parameter. In the undriven case, i.e. for $F(t)=0$, the properties of the Krawtchouk quantum chain are well known and
have been extensively investigated, especially in the context of coherent quantum transport for its capability of realizing perfect quantum state transfer and mirror inversion \cite{K1,K2,K3}. At $F=0$, the energies of $\hat{H}$ are equally-spaced and given by $E_n=\nu (N-2n)$ ($n=0,1,2,...,N$), whereas the corresponding eigenfunctions are expressed in terms of the Krawtchouk polynomials \cite{K0,K2}. Equal energy spacing provides exact periodic wave packet reconstruction with periodicity $T_R=\pi/ \nu$, whereas mirror inversion occurs at the time $T_R/2$.  An interesting property of the Krawtchouk quantum chain is that perfect wave packet reconstruction persists  when a dc force is applied to the particle: in this case the particle undergoes periodic Bloch oscillations on the inhomogeneous lattice \cite{K5}. Here we wish to study the  Krawtchouk quantum chain for a general time-dependent force $F(t)$, and in particular for ac forcing. To this aim, it is worth establishing a correspondence between the hopping motion of a forced quantum particle on the  Krawtchouk chain, expressed by Eqs.(2) and (3),  and the dynamics in Fock space of $(N+1)$ non-interacting bosons trapped in a driven double-well potential, the so-called driven bosonic junction; see Fig.1(a). The Hamiltonian of a non-interacting bosonic field in a double-well potential driven by a time-dependent force is described by the second-quantization Hamiltonian
\begin{equation}
\hat{H}_{B}(t)=-\nu \left( \hat{a}_{L}^{\dag} \hat{a}_R+\hat{a}_{R}^{\dag} \hat{a}_L \right) +\frac{F(t)}{2} (\hat{n}_L-\hat{n}_R)
\end{equation}
  where R (right) and L (left) are the two well sites, $\hat{a}_{k}^{\dag}$
and $\hat{a}_k$ are the bosonic creation and annihilation operators
and $\hat{n}_k =\hat{a}^{\dag}_{k}\hat{a}_k$ the corresponding particle number
operator in the two wells ($k=L,R$), $v$ is the hopping
(tunneling) rate between the two wells, and $F(t)$ is the external
time-dependent force. The total number of bosons $\hat{N}=\hat{n}_L+\hat{n}_R$ is a conserved
quantity and the dimension of the Hilbert space
is $(N + 1)$.  The Hamiltonian (4) considered in this work is obtained from the general bosonic junction model \cite{Bos1,Bos2,Bos3} in the limit of non-interacting particles. 
Note that, since we neglect particle interaction and $\hat{H}_B$ is quadratic in the field operators, the Heisenberg equations of motion for the bosonic operators $\hat{a}^{\dag}_{L,R}$ are linear equations and read
 \begin{eqnarray}
 i \frac{d \hat{a}^{\dag}_{L}}{dt} & = & \left[ \hat{a}^{\dag}_{L}, \hat{H}_B \right]  =  \nu \hat{a}^{\dag}_{R}-\frac{F(t)}{2} \hat{a}^{\dag}_{L} \\
 i \frac{d \hat{a}^{\dag}_{R}}{dt} & = & \left[ \hat{a}^{\dag}_{R}, \hat{H}_B \right]  =  \nu \hat{a}^{\dag}_{L}+\frac{F(t)}{2} \hat{a}^{\dag}_{R}.
 \end{eqnarray} 
Equations (5) and (6) can be formally solved, yielding
 \begin{equation}
 \left(
 \begin{array}{c}
 \hat{a}^{\dag}_L (t) \\
 \hat{a}^{\dag}_R (t)
 \end{array}
 \right)=
 \left(
 \begin{array}{cc}
 S_{11}(t) & S_{12}(t) \\
 S_{21}(t) & S_{22}(t)
 \end{array}
 \right) \times 
 \left(
 \begin{array}{c}
 \hat{a}^{\dag}_L (0) \\
 \hat{a}^{\dag}_R (0)
 \end{array}
 \right)
 \end{equation}
 where the unitary $ 2 \times 2$ matrix $S(t)$ is the propagator of the linear system. Note that the propagator $S(t)$ is the same that one would obtain by replacing  in Eqs.(5) and (6) the operators $\hat{a}^{\dag}_{L,R}$ by $c$-numbers, i.e. $S(t)$ is the propagator of a driven two-level system \cite{Grifoni} that describes {\it single-partilce} tunneling in the double-well of Fig.1(a).
The equivalence between the bosonic junction model, defined by the second-quantization Hamiltonian (4), and the 
single-particle ac-driven Krawtchouk quantum chain, defined by Eqs.(2) and (3),  can be readily proven by 
looking at the dynamics of the bosonic field in Fock space with constant particle
number $N$ \cite{Bos2}. In fact, let us expand the state vector $|\psi(t) \rangle_B$ of the bosonic field in Fock space as 
\begin{equation}
| \psi(t) \rangle_B= \sum_{n=0}^{N}\frac{c_n(t) \exp \left[ i(N/2) \int_0^t dt' F(t') \right]}{\sqrt{n! (N-n)!}} \hat{a}^{\dag \; n}_{L} \hat{a}^{\dag \; N-n}_{R} | 0 \rangle
\end{equation}
 where $c_n(t)$ is the amplitude probability to find $n$ bosons in the left well and the other $(N-n)$ bosons in the right well [see Fig.1(a)], and an additional time-dependent phase term has been added to $c_n(t)$ for the sake of convenience.  Substitution of Eqs.(4) and (8) into the Schr\"{o}dinger equation $ i \partial_t | \psi(t) \rangle_B= \hat{H}_B | \psi(t) \rangle_B$ yields for the amplitudes $c_n(t)$ the same evolution equations as defined by Eqs.(2) and (3) (see, for instance, \cite{Bos2}).

\section{Quasi-energy spectrum, fractional dynamic localization and mirror inversion}

The equivalence between the forced Krawtchouk quantum chain and the bosonic junction model established in the previous section enables to derive a closed-form solution to the time-dependent linear system (2) for an arbitrary forcing $F(t)$. In particular, the conditions for perfect wave packet reconstruction, fractional DL and mirror inversion can be readily obtained from the associated two-level system discussed in the previous section. To this aim, let us consider the initial quantum state $|\psi(0) \rangle_B$ of the bosonic field 
\begin{equation}
| \psi(0) \rangle_B= \sum_{n=0}^{N}\frac{c_n(0)}{\sqrt{n! (N-n)!}} \hat{a}^{\dag \; n}_{L} \hat{a}^{\dag \; N-n}_{R} | 0 \rangle.
\end{equation}
The quantum state $|\psi(t) \rangle_B$ at a successive time $t$ is obtained from Eq.(9) after the replacement
\begin{equation}
\hat{a}^{\dag}_L \rightarrow S_{11}(t) \hat{a}^{\dag}_L+S_{12}(t) \hat{a}^{\dag}_R \; \;\;,\;\;\;\;\;\;  \hat{a}^{\dag}_R \rightarrow S_{21}(t) \hat{a}^{\dag}_L+S_{22}(t) \hat{a}^{\dag}_R
\end{equation}
i.e. one has
\begin{equation}
| \psi(t) \rangle_B= \sum_{n=0}^{N}\frac{c_n(0)}{\sqrt{n! (N-n)!}} \left( S_{11}(t) \hat{a}^{\dag}_{L} +S_{12}(t) \hat{a}^{\dag}_R \right)^n 
 \left( S_{21}(t) \hat{a}^{\dag }_{L} +S_{22}(t) \hat{a}^{\dag}_R \right)^{N-n} | 0 \rangle
\end{equation}
where the unitary matrix $S(t)$ is determined from the analysis of the driven two-level system discussed in the previous section [see Eqs.(5-7)]. A comparison of Eqs.(8) and (11) then yields
\begin{equation}
 \sum_{n=0}^{N}\frac{c_n(t) }{\sqrt{n! (N-n)!}} \hat{a}^{\dag \; n}_{L} \hat{a}^{\dag \; N-n}_{R} | 0 \rangle = \exp \left[ -i(N/2) \int_0^t dt' F(t') \right] \sum_{n=0}^{N}\frac{c_n(0)}{\sqrt{n! (N-n)!}} \left( S_{11}(t) \hat{a}^{\dag }_{L} +S_{12}(t) \hat{a}^{\dag}_R \right)^n 
 \left( S_{21}(t) \hat{a}^{\dag }_{L} +S_{22}(t) \hat{a}^{\dag}_R \right)^{N-n} | 0 \rangle
\end{equation}
which can be formally solved for $c_n(t)$,  obtaining
\begin{equation}
c_n(t)=\sum_{l=0}^{N} \mathcal{U}_{n,l}(t) c_{l}(0)
\end{equation}
where the propagator $\mathcal{U}(t)$ is given by
\begin{equation}
\mathcal{U}_{n,l}(t)=\frac{ \exp \left[ -i(N/2) \int_0^t dt' F(t') \right] }{\sqrt{n! l! (N-n)!(N-l)!}} 
 \langle 0 | \hat{a}_L^n \hat{a}_{R}^{N-n} 
\left( S_{11}(t) \hat{a}^{\dag }_{L} +S_{12}(t) \hat{a}^{\dag}_R \right)^l 
 \left( S_{21}(t) \hat{a}^{\dag }_{L} +S_{22}(t) \hat{a}^{\dag}_R \right)^{N-l} | 0 \rangle
\end{equation}
Using the binomial expansion and commutation relations for the bosonic operators, after some straightforward calculations the following final form for the propagator $\mathcal{U}(t)$ can be derived
\begin{equation}
\mathcal{U}_{n,l}(t)=\exp \left[ -i(N/2) \int_0^t dt' F(t') \right]
\sqrt{\frac{n!(N-n)!}{l!(N-l)!}} \sum_{k=k_1(n,l)}^{k_2(n,l)} 
\left(
\begin{array}{c}
l \\
k
\end{array}
\right)
\left(
\begin{array}{c}
N-l \\
n-k
\end{array}
\right)
S_{11}^{k} (t) S_{21}^{n-k}(t) S_{12}^{l-k}(t)S_{22}^{N-l-n+k}(t)
\end{equation}
where we have set 
\begin{eqnarray}
k_1(n,l) & = & {\rm max} (0,n-N+l) \\
k_2(n,l) & = & \left\{
\begin{array}{cc}
n & {\rm for} \;\;\;\; 2l \geq N \\
{\rm {min}}(n,l)&  \;\;\;\;  {\rm for} \;\;\;\; 2l < N
\end{array}
\right.
\end{eqnarray}
Equation (15), which is the main result of this section, shows that the propagator $\mathcal{U}(t)$ for the forced Krawtchouk quantum chain can be retrieved from the simpler propagator $S(t)$ of the associated two-level problem Eqs.(5-7), regardless of the number $(N+1)$ of lattice sites. In particular, for a periodic ac driving force with frequency $\omega=2 \pi /T$ [i.e. $F(t+T)=F(t)$], Floquet theory applies and the quasi energy spectrum of the  ac-driven Krawtchouk quantum chain can be obtained from the Floquet exponents of the propagator $\mathcal{U}$ at time $t=T$. 
Much of the propagation properties of the  ac-driven Krawtchouk quantum chain are determined by the properties of the two-level system Eqs.(5) and (6). In fact, let us indicate by $\mu$ and $-\mu$ the two quasi energies of the two-level system, i.e. the Floquet exponents of the propagator $\mathcal{S}(t=T)$ \cite{Grifoni}. According to Eq.(15) and Floquet theory, the propagator $\mathcal{U}(t)$ is composed by only {\it two} ladders of frequencies, $n \omega \pm \mu$ ($n=0, \pm 1, \pm2, ...$), which are shifted $2 \mu$ apart each  other. Hence the following properties hold rather generally:\\
(i) If $\mu=0$, perfect wave packet reconstruction is observed at times $t=T,2T, 3T, ...$ and exact quasi energy band collapse is attained. This regime basically corresponds to the ordinary DL regime \cite{Dunlap,Holthaus}, where periodic  wave packet reconstruction occurs at integer multiplies of the driving period $T$. Note that the condition $\mu=0$ corresponds to the coherent destruction of tunneling (CDT) for the two-level system of Eqs.(15-17), i.e. for the double well of Fig.1(a) \cite{Grifoni}. Hence the ac-driven Krawtchouk quantum chain provides a rather exceptional example of inhomogeneous chain where multilevel CDT can be realized  in an exact way \cite{LonghiPRB2008}.\\
(ii) If 
\begin{equation}
\frac{2 \mu}{\omega}=\frac{Q}{M} 
\end{equation}
 with $Q$ and $M$ irreducible integers, perfect wave packet reconstruction is obtained at times $t$ multiplies that $T_M=MT=2 \pi M / \omega$. This regime can be referred to as {\it fractional} DL, because it corresponds to the ordinary DL regime but with a driving force with a fictitious frequency $ \omega /M$. As it will be discussed in the next section, in such a regime the $(N+1)$ quasi energies of the  Krawtchouk  quantum chain undergo $M$ partial collapses at values which are 
integer multiplies than $\omega/M$. It should be noted that, since the quasi energy $\mu$ is a continuous function of the amplitude of the driving force, it follows that fractional DL localization occurs for {\it almost every value} of the driving amplitude. In the framework of the general theory of quantum recurrence in Hamiltonian systems with a discrete energy spectrum \cite{Loinger,periodic}, fractional DL can be viewed as an {\it exact} realization of quantum recurrence.\\
(iii) If $ 2 \mu / \omega$ is an irrational number, perfect wave packet reconstruction is not realized at any time and the dynamics is quasi-periodic. According to the general recurrence theorem of ac-driven quantum systems with a discrete quasi energy spectrum \cite{periodic}, any arbitrary initial quantum state can nevertheless reconstruct itself with any desired degree of approximation after some (possibly long) time. \\
(iv) If at time $t=\tau$ the propagator $S(\tau)$ of the two-level system has the form
\begin{equation}
S(\tau)=\left(
\begin{array}{cc}
0 & \exp(i \varphi) \\
\exp(-i \varphi) & 0
\end{array}
\right)
\end{equation}
where $\varphi$ is an arbitrary phase term, i.e. if after a time $\tau$ the propagation flips one particle in Fig.1(a)  from one well to the other one, then the propagator $\mathcal{U}(\tau)$ of the Krawtchouk quantum chain realizes mirror inversion. In fact, if Eq.(19) holds, from Eq.(11) one obtains
\begin{equation}
| \psi(\tau) \rangle_B= \sum_{n=0}^{N}\frac{c_n(0)}{\sqrt{n! (N-n)!}} \exp(2 i \varphi n-iN \varphi ) \hat{a}^{\dag \; n}_{R}
 \hat{a}^{\dag \; N-n }_{L} =\sum_{n=0}^{N}\frac{c_n(0)}{\sqrt{n! (N-n)!}} \exp(-2 i \varphi n+iN \varphi ) \hat{a}^{\dag \; n}_{L}
 \hat{a}^{\dag \; N-n }_{R} 
 \end{equation}
A comparison of Eqs.(8) and (20) then yields
\begin{equation}
c_n(\tau)=c_{N-n}(0) \exp \left[ i \varphi (N -2  n) -i (N/2) \int_0^{\tau} dt F(t) \right]
\end{equation}
i.e. mirror inversion ($n \rightarrow N-n$) is realized, apart from an additional phase term.\\

\section{Numerical results}
We apply the general theory discussed in the previous section to the case of a sinusoidal driving force
\begin{equation}
F(t)=F_0 \cos (\omega t)
\end{equation}
and assume, without loss of generality, $\nu=1$ \cite{note2}. The quasi energies $\pm \mu$ of the associated two-level system [Eqs.(5-7)] versus the normalized forcing amplitude $\Gamma=F_0 / \omega$ have to be computed in the general case only numerically; in the large frequency limit $\omega \gg 1$ they are approximated by the simple relation \cite{Grifoni}
\begin{equation}
\mu \simeq J_0(\Gamma),
\end{equation}
where $J_0$ is the Bessel function of first kind and zero order. As an example, in Fig.1(b) we show the numerically-computed quasi energies versus $\Gamma$  of the two-level system for $\omega=3$. Note that quasi energy crossing, corresponding to CDT and $\mu=0$, is attained at $\Gamma \equiv \Gamma_0 \simeq 2.261$. The corresponding quasi energy spectrum for the  Krawtchouk quantum chain comprising $(N+1)=41$ lattice sites is shown in Fig.2. The quasi energy spectrum versus the normalized amplitude $\Gamma$ shows a characteristic and fine pattern, which is basically determined by the circumstance that, as $\Gamma$ is varied, the ratio $2 \mu(\Gamma) / \omega$ can be either a rational or an irrational number \cite{noteHous}. At $\Gamma=\Gamma_0 \simeq 2.261$, {\it full} collapse of the $(N+1)=41$ quasi energies is clearly observed, corresponding to the ordinary DL regime. However, a {\it partial} collapse of quasi energies can be clearly observed at many values of the  normalized driving amplitude, according to the general analysis presented in the previous section. For example, in Fig.2 the vertical arrows highlight partial quasi energy collapses occurring at four values of $\Gamma$, namely at $\Gamma \simeq 0.496, 0.84, 1.3172$ and $1.5476$, corresponding to $5,2,3$ and 4 partial collapses, respectively. Note that, at such values of the normalized forcing amplitude, the ratio $ 2 \mu / \omega$ is a rational number, namely one has $2 \mu / \omega=Q/M=3/5, 1/2, 1/3$ and $1/4$, respectively, i.e. $M=5,2,3$ and $4$. According to the analysis presented in the previous section, at such values of the normalized driving amplitude exact wave packet reconstruction, corresponding to fractional DL, should be observed at times multiplies than $MT= 2 M \pi/3$. This is clearly shown in Fig.3. In the left panels of Figs.3(a-d) we show the numerically-computed evolution of the occupation probabilities $|c_n(t)|^2$ for the initial condition $c_n(0)=\delta_{n,0}$ and for the four values of $\Gamma$ corresponding to fractional DL indicated by the vertical arrows in Fig.2. In the right panels of the figures, the detailed evolution of $|c_0(t)|^2$ and $|c_N(t)|^2$ is also shown. The figures clearly show that wave packet reconstruction is attained at time $MT$ (and integer multiplies). Note also that, in case of Fig.3(a) and (c), i.e. for an odd value of $M$, mirror inversion is also observed at time $TM/2$. This is due to the fact that, according to the property (iv) discussed in the previous section, for odd values of $M$ the two-level system flips the two states at the time $TM/2$.
In Figure 4 we show, for comparison,  the evolution of the occupation probabilities in the DL regime [Fig.4(a), $\Gamma=\Gamma_0=2.261$], and in the quasi-periodic regime [Fig.4(b), $\Gamma=0.9965$], where $ 2 \mu / \omega=1/ \sqrt{5}$ is an irrational number. In the latter case perfect wave packet reconstruction is not observed, however quantum recurrence can be clearly seen over the observation time.


\section{Conclusion and discussion}

In this paper we have theoretically investigated quantum recurrence and dynamic localization phenomena in the ac-driven Krawtchouk quantum chain. We have shown that the hopping dynamics in the quantum chain with an arbitrary number of $N+1$ sites can be mapped into the tunneling problem of $N$ non interacting bosons in an ac-driven double well system. Such an equivalence enables to establish a few important properties of quantum transport in the  ac-driven Krawtchouk quantum chain and to predict a new type of dynamic localization. In particular, we have shown that quantum recurrence, which is predicted by the general quantum recurrence theorem, is {\it exact} for the  Krawtchouk quantum chain in a dense range of the driving amplitude and related to the existence of so-called fractional dynamic localization. As opposed to ordinary DL, at which full collapse of the quasi energy spectrum occurs, fractional DL corresponds  to a set of partial quasi-energy collapses. Moreover, like for the undriven  Krawtchouk quantum chain, it has been shown that mirror inversion can be also observed in fractional DL under certain conditions. The phenomenon of fractional DL, which is the main result of the present study, could be experimentally demonstrated is spin chains using nuclear magnetic resonance \cite{nuc} or in photonic transport using evanescently-coupled optical waveguides arrays with engineered couplings \cite{K4}. In the latter system, the ac driving filed can be simply mimicked by periodical axis bending of the waveguide structure \cite{optics,optics2,pisello}.\par  It is envisaged that our results could be of interest in the field of coherent quantum transport and control, where transferring quantum states efficiently between distant nodes of an information processing circuit
is of major importance for scalable quantum computing. For example, exploitation of fractional DL could be important for controlling and tune the coherent transfer time for mirror inversion: while in the undriven (static) Krawtchouk quantum chain this time is fixed by the hopping rate $\nu$ of the engineered Hamiltonian, in the ac-driven chain  the transfer time is determined by the frequency of the external field and can be thus properly tuned.


\begin{thebibliography}{00}
\bibitem{note2}
The $\nu=1$ case is obtained after proper normalization of the time variable $t$.
\end{thebibliography}
\end{document}